# A Forty Year Journey[1]

Reinhard Genzel

Max-Planck Institute for Extraterrestrial Physics, Garching, Germany
Departments of Physics & Astronomy, University of California, Berkeley, USA

## 1. Prologue

A *'black hole'* (e.g. Wheeler 1968) conceptually is a region of space-time where gravity is so strong that within its event horizon neither particles with mass, nor even electromagnetic radiation (massless photons), can escape from it. Based on Newton's theory of gravity, Rev. John Michell (in 1784) and Pierre-Simon Laplace (in 1795) were the first to note that a sufficiently compact, massive star may have a surface escape velocity exceeding the speed of light. Such an object would thus be 'dark' or invisible. A proper mathematical treatment of this remarkable proposition had to await Albert Einstein's theory of General Relativity in 1915/1916 (henceforth GR). Karl Schwarzschild's (1916) solution of the vacuum field equations in spherical symmetry demonstrated the existence of a characteristic *event horizon* of a mass M, the ***Schwarzschild radius*** $R_s=2GM/c^2$, within which no communication is possible with external observers. It is a 'one way door'. Roy Kerr (1963) generalized this solution to spinning black holes. However, these solutions refer to configurations with sufficiently high symmetry, so that Einstein's equations can be solved analytically, and there was doubt about whether such cases were typical. Roger Penrose, one of the other recipients of this year's Nobel Prize, dropped the assumption of spherical symmetry, and analyzed the problem topologically (Penrose 1963, 1965). Using the key concept of 'trapped surfaces' he showed that any arbitrarily shaped surface with a radius less than the Schwarzschild radius is a trapped surface, and the radial direction becomes time-like as one passes through the horizon. Any observer is then inexorably pulled towards the center where time ends. All the matter that forms the black hole resides at this single moment in time, the singularity.

From considerations of the information content of black holes, there is significant tension between the predictions of GR and general concepts of quantum theory (e.g. Susskind 1995, Maldacena 1998, Bousso 2002). It is likely that a proper quantum theory of gravity will modify the concepts of GR on scales comparable to or smaller than the Planck length, $l_{Pl}\sim 1.6 \times 10^{-33}$ cm, remove the concept of the central singularity, and potentially challenge the interpretation of the GR event horizon (Almheiri et al. 2013).

But are these bizarre objects of GR actually realized in Nature?

## 2. Overture: X-ray Binaries & Quasars

Astronomical evidence for the existence of black holes started to emerge sixty years ago with the discovery of variable X-ray emitting binaries in the Milky Way (Giacconi et al. 1962, Giacconi 2003 (Nobel Lecture 2002)) on the one hand, and of distant luminous 'quasi-stellar-radio-sources/objects' (QSOs, Schmidt 1963) on the other. For about two dozen X-ray binaries, dynamical mass determinations from Doppler spectroscopy of the visible primary star established that the mass of the X-ray emitting secondary is significantly larger than the maximum stable neutron star mass,

---

[1] Nobel Lecture, December 8, 2020

~2.3 solar masses (McClintock & Remillard 2004, Remillard & McClintock 2006, Özel et al. 2010, Rezzolla et al. 2018). The binary X-ray sources thus are excellent candidates for stellar black holes (SBH). They are probably formed when a massive star explodes as a supernova at the end of its fusion lifetime and the compact remnant collapses to a SBH. The measurements of gravitational waves from in-spiraling binaries with LIGO (Abbott et al. 2016a, b, Nobel Prize 2017) have recently provided very strong ,and arguably conclusive evidence for the existence of SBHs.

The luminosities of QSOs often exceed by 3 to 4 orders of magnitude the entire energy output of the Milky Way Galaxy. Furthermore their strong high energy emission in the UV-, X-ray and γ-ray bands, as well as their spectacular relativistic jets, can most plausibly be explained by accretion of matter onto massive black holes (henceforth MBHs, e.g. Lynden-Bell 1969, Shakura & Sunyaev 1973, Blandford 1999, Yuan & Narayan 2014, Blandford, Meier & Readhead 2019). Between 7% (for a non-rotating Schwarzschild hole) and 40% (for a maximally rotating Kerr hole) of the rest energy of an infalling particle can, in principle, be converted to radiation outside the event horizon, one to two orders of magnitude greater than nuclear fusion in stars. To explain powerful QSOs by this mechanism, black hole masses of $10^8$ to $10^9$ solar masses and accretion flows between 0.1 to 10 solar masses per year are required. QSOs are located (without exception) in the nuclei of large, massive galaxies (e.g. Osmer 2004). QSOs represent the most extreme and spectacular among the general nuclear activity of most galaxies.

A conclusive experimental proof of the existence of a SBH or MBH, as defined by GR, requires the ***determination of the gravitational potential on the scale of the event horizon***. This gravitational potential can be inferred from spatially resolved measurements of the motions of test particles (interstellar gas, stars, other black holes, or photons) in close orbit around the black hole (Lynden-Bell & Rees 1971). Until very recently this ambitious test was not feasible. A more modest goal then is to show that the gravitational potential of a galaxy nucleus is dominated by a compact non-stellar mass and that this central mass concentration cannot be anything but a black hole, because all other conceivable configurations either are more extended, are not stable, or produce more light (e.g. Maoz 1995, 1998). Even this test cannot be conducted (yet) in distant QSOs. Lynden-Bell (1969) and Lynden-Bell & Rees (1971) proposed that MBHs might be common in most galaxies (although in a low state of accretion). If so dynamical tests are feasible in nearby galaxy nuclei, including the Center of our Milky Way.

Over the past fifty years, since these seminal papers, increasingly solid ***evidence for central 'dark' (i.e. non-stellar) mass concentrations*** has emerged for about one hundred galaxies (e.g. Kormendy 2004, Gültekin et al. 2009, Kormendy & Ho 2013, McConnell & Ma 2013, Saglia et al. 2016, Greene et al. 2016), from optical/infrared imaging and spectroscopy on the Hubble Space Telescope (HST) and large ground-based telescopes, as well as from Very Long Baseline radio Interferometry (VLBI). Further evidence comes from relativistically broadened, redshifted iron Kα line emission in nearby Seyfert galaxies (e.g. Tanaka et al. 1995, Nandra et al. 1997, Fabian et al. 2000). In external galaxies the most compelling case that such a dark mass concentration cannot just be a dense nuclear cluster of white dwarfs, neutron stars and perhaps stellar black holes emerged in the mid-1990s from spectacular VLBI observations of the nucleus of NGC 4258, a mildly active galaxy at a distance of 7 Mpc (Miyoshi et al. 1995, Moran 2008). The VLBI observations show that the galaxy nucleus contains a thin, slightly warped disk of $H_2O$ masers (viewed almost edge on) in Keplerian rotation around an unresolved mass of 40 million solar masses. The inferred density of this mass exceeds a few $10^9$ solar masses $pc^{-3}$ and thus cannot be a long-lived cluster of 'dark' astrophysical objects of the type mentioned above (Maoz 1995). As we will discuss below, the Galactic Center provides a yet more compelling case.

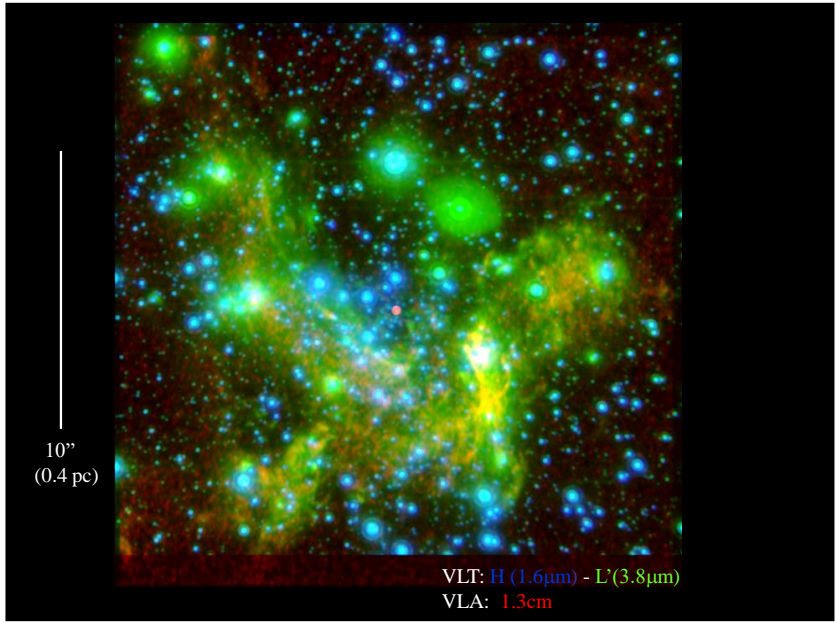

Figure 1. Near-infrared/radio, color-composite image of the central light years of the Galactic Center. The blue and green colors represent the 1.6 and 3.8μm broad-band near-infrared emission, at the diffraction limit (~0.05") of the 8m Very Large Telescope (VLT) of the European Southern Observatory (ESO), and taken with the 'NACO' AO-camera and an infrared wave-front sensor (adapted from Genzel et al. 2003a). Similar work has been carried out at the 10m Keck telescope (Ghez et al. 2003, 2005). The red color image is the 1.3cm radio continuum emission taken with the Very Large Array (VLA) of the US National Radio Astronomy Observatory (NRAO). The red dot in the center of the image is the compact, non-thermal radio source SgrA*. Many of the bright blue stars are young, massive O/B- and Wolf-Rayet (WR) stars that have formed recently. Other bright stars are giants and asymptotic giant branch stars in the old nuclear star cluster. The extended streamers/wisps of 3.8μm emission and radio emission are dusty filaments of ionized gas orbiting in the central light years (adapted from Genzel, Eisenhauer & Gillessen 2010).

## 3. Scherzo: SgrA* & Gas Motions

The central light years of our Galaxy contain a dense and luminous star cluster, as well as several components of neutral, ionized and extremely hot gas (Figure 1, Genzel & Townes 1987, Genzel, Hollenbach & Townes 1994, Melia & Falcke 2001, Genzel, Eisenhauer & Gillessen 2010, Morris, Meyer & Ghez 2012, Reid 2013). Compared to the distant QSOs, the Galactic Center is 'just around the corner' ($R_0$=8.25 kilo-parsecs (kpc), 27,000 light years). High-resolution observations of the Milky Way nucleus thus offer the unique opportunity of carrying out a stringent test of the MBH-paradigm deep within its gravitational 'sphere of influence' where gravity is dominated by the central mass (R<1-3 pc). Since the Center of the Milky Way is highly obscured by interstellar dust particles in the plane of the Galactic disk, observations in the visible part of the electromagnetic spectrum are not possible. The veil of dust, however, becomes transparent at longer wavelengths (the infrared, microwave and radio bands), as well as at shorter wavelengths (hard X-ray and γ-ray bands), where observations of the Galactic Center thus become feasible (Oort 1977).

The stellar density in the nuclear cluster increases inward from a scale of tens of parsecs to within the central 0.04 parsec (Becklin & Neugebauer 1968, Genzel et al. 2003a). At its center is a **_very compact radio source, Sgr A*_** (Fig. 1, Balick & Brown, 1974, Lo et al. 1985, Backer et al. 1993). Millimeter inter-continental Very Long Baseline Interferometry (VLBI) observations have established that its intrinsic radius is a mere 20-50 micro-arcseconds (µas) (Figure 2, 2-5 $R_S$ for a $4 \times 10^6$ $M_\odot$ MBH; Krichbaum et al. 1993, Bower et al. 2004, Shen et al. 2005, Doeleman et al. 2008, Johnson et al. 2015, Lu et al. 2014, 2018, Issaoun et al. 2019). Sgr A* thus is the prime candidate for the location and immediate environment of a possible MBH.

VLBI observations also have set an upper limit of about 0.6 km/s and 1 km/s to the motion of SgrA* itself, along and perpendicular to the plane of the Milky Way, respectively (Reid & Brunthaler 2004, 2020). When compared to the two orders of magnitude greater velocities of the stars in the immediate vicinity of SgrA* (see below), this demonstrates that the radio source must indeed be massive, with simulations giving a lower limit to the mass of SgrA* of ~$10^5$ $M_\odot$ (Chatterjee, Hernquist & Loeb 2002, but see Tremaine, Kocsis & Loeb 2021).

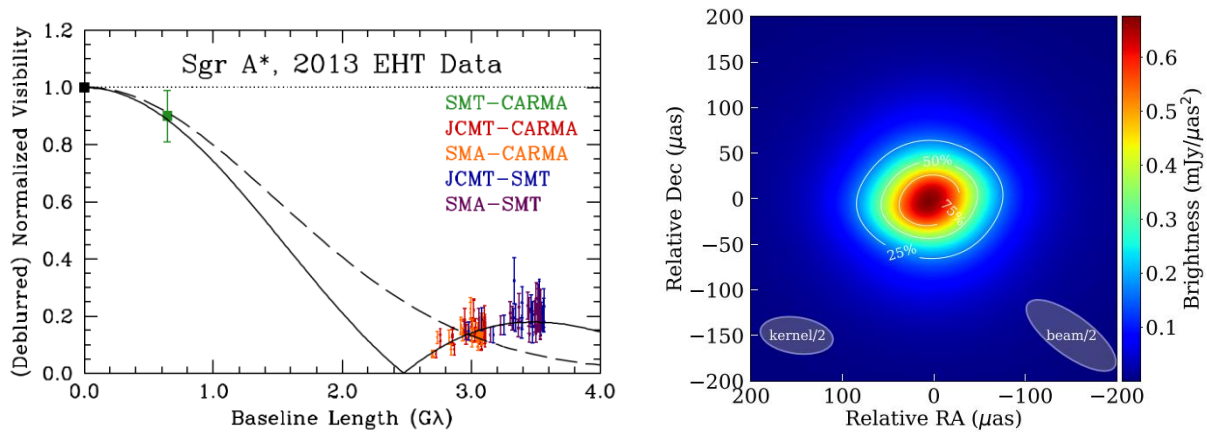

Figure 2. Total-intensity mm-VLBI of Sgr A∗. Left: Normalized, de-blurred visibilities at 1.3mm taken with the Event Horizon Telescope are shown as a function of baseline length; errors are ±1σ. The dashed line shows the best-fit circular Gaussian (FWHM: 52 µas). An annulus of uniform intensity (inner diameter: 21 µas, outer diameter: 97 µas), shown with a solid line, is perhaps the most plausible model that is consistent with the data (adapted from Figure S5 in Johnson et al. 2015, Supplement). Right: 3mm Global mm-VLBI image of SgrA*, after removal of the scattering screen. The reconstructed image has an intrinsic Gaussian source diameter of $\theta_{maj}$ = 120 ± 34 µas and $\theta_{min}$ = 100 ± 18 µas. The ellipses at the bottom indicate half the size of the scatter-broadening kernel ($\theta_{maj}$ = 159.9 µas, $\theta_{min}$ = 79.5 µas, PA = 81.9°) and of the observing beam (adapted from Figure 5 in Issaoun et al. 2019).

The first ***dynamical evidence for the presence of a non-stellar mass concentration of 2-4 million times the mass of the Sun ($M_\odot$)*** and plausibly centered on or near SgrA* came from infrared imaging spectroscopy of interstellar gas clouds, carried out by Charles Townes' group in Berkeley[2] (Wollman et al. 1977, Lacy et al. 1980, Serabyn & Lacy 1985, Crawford et al. 1985). In their 1985 Nature paper Crawford et al. summarized the then available evidence on the mass distribution obtained from the infrared and submillimeter spectroscopy that traced the ionized and neutral gas components. They concluded that "…the measurements fit a point mass of ~$4 \times 10^6$ $M_\odot$ but are also consistent with a cluster where stellar density decreases with radius (*R*) at least as fast as $R^{-2.7}$, or a combination of a point mass and a stellar cluster….." However, many considered this

---
[2] I had joined Townes' group in 1980 as a Miller Postdoctoral Fellow, and then became Associate Professor in the Physics Department in 1981

## 4. Escursione: Ever sharper, Ever deeper

The most critical aspect in testing the MBH paradigm obviously lies in the ability of sensitive, very high angular resolution observations. The Schwarzschild radius of a 4 million solar mass black hole at the Galactic Center subtends a mere $10^{-5}$ arc-seconds, or 10μas[3].

In the *radio and millimeter* bands such high resolution can be obtained from VLBI. Starting in the 1980s, ever higher resolution VLBI measurements showed that the radio size of SgrA* decreases with decreasing wavelength, owing due to scattering by intervening electrons between SgrA* and the Earth (Shen et al. 2005, Bower et al. 2006). Measuring the intrinsic size of the source and imaging its two dimensional distribution requires short millimeter VLBI observations, which are technically very challenging (Event Horizon Telescope Project in the USA: Doeleman 2010, Black Hole Cam Project in Europe: Goddi et al. 2017)[4].

For high-resolution *infrared* imaging from the ground, an important technical hurdle is the correction of the distortions of an incoming electromagnetic wave by the turbulent, refractive Earth atmosphere. In the optical/near-infrared wave-band the atmosphere distorts the incoming electromagnetic waves on time scales of milli-seconds and smears out long-exposure images to a diameter of more than an order of magnitude greater than the diffraction limited resolution of large ground-based telescopes. The enormous progress in testing the MBH paradigm in the Galactic Center carried out by our group at MPE (at the telescopes of the European Southern Observatory in Chile), and by Andrea Ghez and her collaborators (at the Keck telescopes in Hawaii), described in the following sections, largely rests on substantial, continuous improvements in the angular resolution, astrometric precision and sensitivity of near-IR imaging and spectroscopy (by factors between one hundred to one hundred thousand over three decades).

From the early 1990s onward, short exposure imaging with new infrared imaging detectors were made possible with 'speckle imaging', resulting in diffraction limited resolution (0.05-0.1") near-infrared stellar images (Sibille, Chelli & Léna 1979, Christou 1991, Hofmann et al. 1993, Matthews & Soifer 1994). Because of the short exposures and detector noise, speckle imaging is not able to go very deep. In the early 1990s, 'adaptive optics' techniques (AO: correcting the wave distortions on-line) became available (Rousset et al. 1990, Lena 1991, Tyson & Wizinowich 1992), with upgraded imaging cameras (Lenzen et al. 2003, Lenzen & Hofmann 1995), which have since allowed increasingly precise high-resolution near-infrared observations with the currently largest (10 m diameter) ground-based telescopes. If bright natural guide stars near the science target are not available, laser guide star beacons can be employed for AO corrections (Max et al. 1997, Rabien et al. 1998, Bonaccini-Calia et al. 2006, Ghez et al. 2005). Increasingly powerful integral field spectrometers (IFUs) coupled with AO have opened up deep imaging spectroscopy near the diffraction limit (Weitzel et al. 1994, Eisenhauer et al. 2003b, Larkin et al. 2006). The most recent step forward in the capability of the impressive record of instrumental innovation brought to bear on Galactic Center MBH studies is spatial interferometry, which I discuss separately below (Glindemann et al. 2003, Eisenhauer et al. 2008, 2011, GRAVITY collaboration et al. 2017).

---

[3] 10 μas correspond to about 2cm at the distance of the Moon
[4] https://eventhorizontelescope.org/ , https://blackholecam.org/

## 5. Menuetto: Stellar Motions and Orbits

A more reliable probe of the gravitational field are stellar motions, which started to become available from Doppler spectroscopy of stellar absorption and emission lines in the late 1980s. They broadly confirmed the results obtained in Phase 1 from gas motions (Rieke & Rieke 1988, McGinn et al. 1989, Sellgren et al. 1990, Krabbe et al. 1991, 1995, Haller et al. 1996, Genzel et al. 1996). As described in the last section, the ultimate breakthrough came from the combination of AO techniques with IFU imaging spectroscopy (Eisenhauer et al. 2003b), opening deep near-infrared spectroscopy of thousands of O/B and WR stars and GKM giants (e.g. Trippe et al. 2008, Do et al. 2013, 2018, Feldmeier et al. 2014, Fritz et al. 2016, Habibi et al. 2019).

With diffraction-limited 'speckle' imagery starting in 1991/1992 on the 3.5m New Technology Telescope (NTT) of the European Southern Observatory (ESO) in La Silla/Chile, our group at MPE was able to determine proper motions of stars as close as ~0.1" from SgrA* (Eckart & Genzel 1996, 1997, Genzel et al. 1997). In 1995 Andrea Ghez's group at the University of California, Los Angeles started a similar program with the 10m diameter Keck telescope in Hawaii (Ghez et al. 1998). Both groups independently found that the stellar velocities follow a 'Kepler' law ($v \sim R^{-1/2}$) as a function of distance from SgrA* and reach $\geq 10^3$ km/s within the central light month. Assuming that the mass in the center is the sum of a point mass and an isothermal star cluster, the central mass inferred from projected mass estimators (Bahcall & Tremaine 1981) is ~2.5 million solar masses, for an isotropic velocity distribution (Figure 3), in excellent agreement between the two groups. For more elliptical orbits the inferred mass increases (Bahcall & Tremaine 1981). We now know that the velocity distribution of the innermost stars favors highly elliptical orbits (Schödel et al. 2003, Gillessen et al. 2017), so that the appropriately corrected estimate of M(0) would be $3.5\text{-}4.7 \times 10^6$ $M_\odot$, for R(GC)=8.25 kpc.

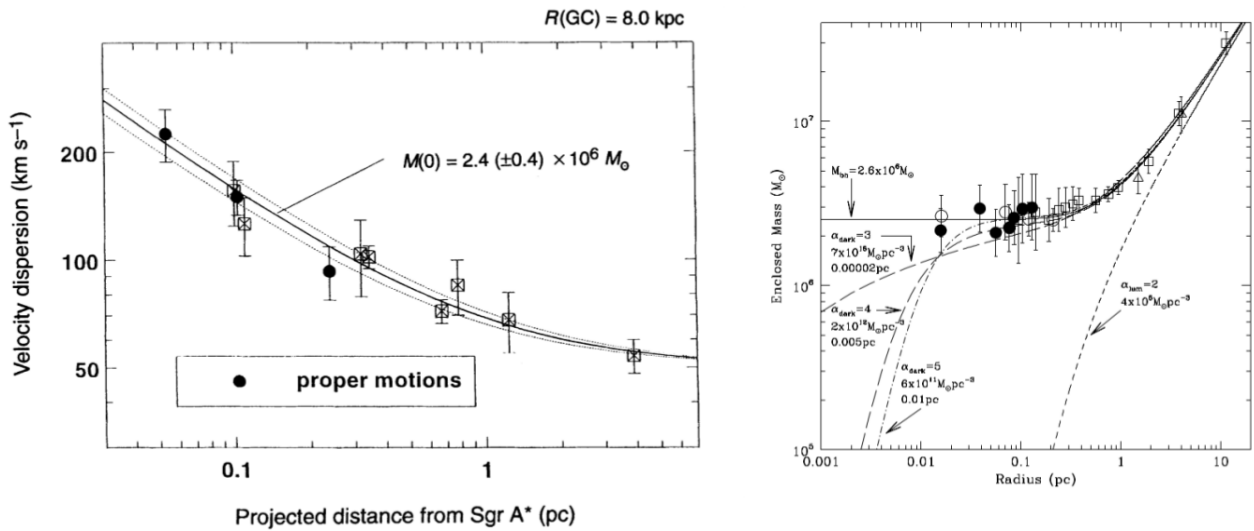

Figure 3. Mass distribution in the central parsec of the Galactic Center after phase 2 (1996/1998). The left graph shows the projected $1^d$ stellar velocity dispersion as a function of projected distance from SgrA*, obtained from proper motions (filled circles) and Doppler velocities (crossed squares) (adapted from Figure 2 of Eckart & Genzel 1996). Each point is derived from averaging the motions of 9-20 stars. The solid curve is a model assuming that the stars move with an isotropic velocity distribution in the potential of a point mass (M(0)) plus an isothermal star cluster of velocity dispersion 50 km/s. The distance of the Galactic Center is assumed to be 8.0 kpc (from Eckart & Genzel 1996). The right graph shows the mass distribution derived from stellar proper motions published by the Keck group in 1998 (Ghez et al. 1998, filled black circles), and compared to the Eckart & Genzel (1996, 1997) proper motions (open circles), the Genzel et al. (1996) stellar radial velocities (squares), and the Guesten et al. (1987) measurement of the rotating gas disk (triangles). From 0.1 to 0.015 pc the enclosed mass appears to be constant with a value of $2.6 \times 10^6$ $M_\odot$. For comparison there are several power law distributions (adapted from Figure 7 of Ghez et al. 1998). ***The agreement between the results of the MPE and UCLA groups is excellent.***

In the next phase, the MPE group moved in 2002 onto ESO's 8.2m Very Large Telescope (VLT) on Paranal, and both groups improved their imagery with adaptive optics and upgraded cameras, improving the astrometry to a few hundred µas in the next decade (Schödel et al. 2002, 2003, 2005, Ghez et al. 2003, 2008, Gillessen et al. 2009a,b, 2017, Meyer et al. 2012, Boehle et al. 2016, Jia et al. 2019). Ghez et al. (2000) detected accelerations for three of the innermost 'S'-stars (subsequently confirmed by Eckart et al. 2002), opening the prospect of much more precise mass determinations from individual orbits, instead of the statistical evaluation through mass estimators.

In 2001/2002 the star S2 (S02) approached SgrA* to 15 mas and made a sharp turn around the radio source during 2002 (Schödel et al. 2002, Ghez et al. 2003). S2/S02 is on a highly elliptical orbit (e=0.88), with a peri-distance of 14 mas (17 light hours or 1400 $R_S$, for M(0)=4.26x$10^6$ $M_\odot$ Figure 4) and an orbital period of a mere 16 years. Ghez et al. (2003, 2005) and Eisenhauer et al. (2003a, 2005) also obtained Doppler velocities and accelerations of S2/S02 and several other orbiting stars, allowing precision measurement of the three dimensional structure of the orbits, as well as the distance to the Galactic Center. Figure 4 shows the data and best fitting GR orbit for S2/S02 in its most recent version (from GRAVITY collaboration et al. 2020a, see below). At the time of writing, the two groups have determined individual orbits for more than 40 stars in the central light month. These orbits show that ***the gravitational potential indeed is dominated by a point mass, whose position is identical within a mas uncertainty with that of the radio source SgrA*** (Plewa et al. 2015, Sakai et al. 2019).

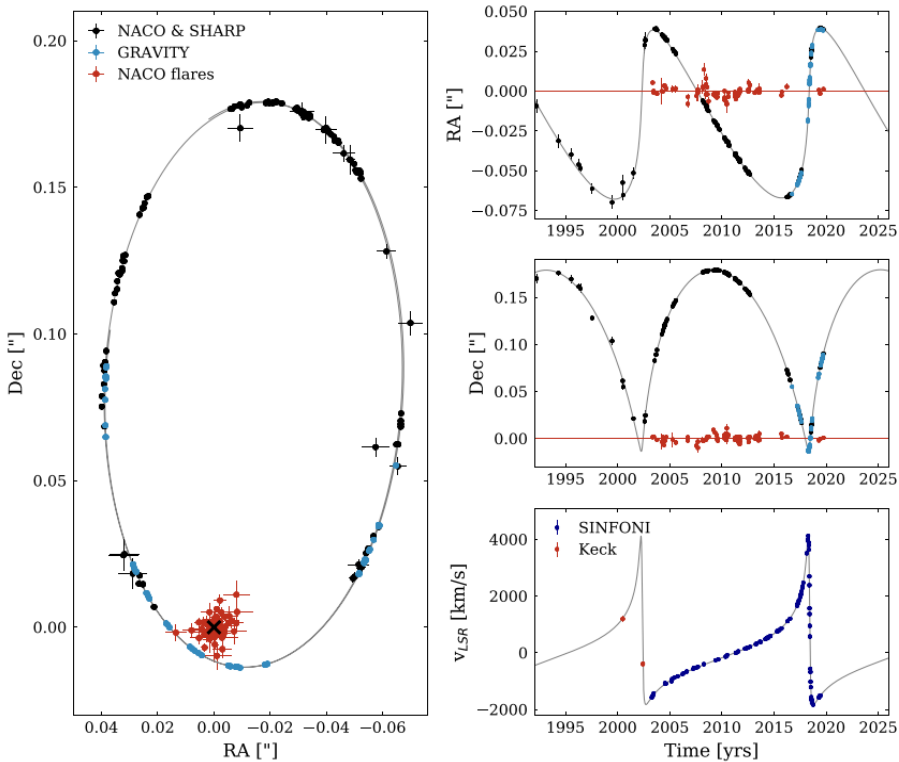

Figure 4. Summary of the MPE-ESO observational results of monitoring the S2-Sgr A* orbit from 1992 to the end of 2019. Left: SHARP (black points with large error bars), NACO (black points), and GRAVITY (blue points) astrometric positions of the star S2, along with the best-fitting GR orbit (grey line). The orbit does not close as a result of the Schwarzschild precession (see text). The mass center is at (0,0), marked by the black cross. All NACO and SHARP points were corrected for a zero-point offset and drift of the reference frame in RA and Dec. The red data points mark the positions of the infrared emission from Sgr A* during bright states, where the separation of S2 and Sgr A* can be directly inferred from differential imaging. Right: RA (top) and Dec (middle) offset of S2 and of the infrared emission from Sgr A* relative to the position of Sgr A* (assumed to be identical with the mass center) (same symbols as in the left panel). Grey is the best-fitting GR-orbit including the Rømer effect (finite speed of light), special relativity, and GR to 'parameterized post-Newtonian' approximation PPN1 (Will 2008). Bottom right: same for the line-of-sight velocity of the star. Position on the sky as a function of time (left) and Doppler velocity (relative to the Local Standard of Rest) as a function of time (right) of the star S2 orbiting the compact radio source SgrA*. Blue filled circles denote data taken with the SINFONI red open circles denote data taken with the Keck telescope as part of the UCLA monitoring project (Do et al. 2019) (adapted from Figure 1 of GRAVITY collaboration 2020a).

At the end of phase 3 (~2017), it is clear that >98% of the 4 million solar mass central mass concentration identified in the first phase is indeed confined to a region <17 light hours around the compact radio source (in a volume a million times smaller than inferred in 1985). The intrinsic size in turn is only a few times the event horizon of that mass. This evidence *eliminates all astrophysically plausible alternatives to a massive black hole*. These include astrophysical clusters of neutron stars, stellar black holes, brown dwarfs and stellar remnants (e.g., Maoz 1995, 1998; Genzel et al. 1997, 2000; Ghez et al. 1998, 2005), and even fermion balls (Viollier, Trautmann & Tupper 1993, Munyaneza, Tsiklauri & Viollier 1998, Ghez et al. 2005; Genzel, Eisenhauer & Gillessen 2010). Clusters of a very large number of mini-black holes and boson balls (Torres, Capozziello & Lambiase 2000; Schunck & Mielke 2003; Liebling & Palenzuela 2012) are harder to exclude. The former have a large relaxation and collapse time, the latter have no hard surfaces that could exclude them from luminosity arguments (Broderick, Loeb & Narayan 2009), and they are consistent with the dynamical mass and size constraints. However, such a boson 'star' would be unstable to collapse to a MBH when continuously accreting baryons (as in the Galactic Center), and it is very unclear how it could have formed. *Under the assumption of the validity of General Relativity the Galactic Center thus provides the best quantitative evidence that MBHs do indeed exist.*

## 6. Rondo Allegretto: Testing General Relativity with SgrA*

At peri-passage S2 moves at v~7650 km/s and $\beta=v/c\sim0.026$ so that the first order Post-Newtonian effects of GR (PPN1: $\sim\beta^2\sim6.5\times10^{-4}$, Will 2008), namely the gravitational redshift and the Schwarzschild in plane orbital precession can be realistically detected in the spectra and the astrometry of the star near peri-center. Knowing that S2 would return in 2018 for its next peri-passage, we proposed in 2005 to ESO to build a novel near-infrared beam combiner instrument (GRAVITY) combining the light of all four 8m telescopes of the VLT (Eisenhauer et al. 2008, Paumard et al. 2008). GRAVITY would improve the angular resolution and astrometry by more than an order of magnitude and thus reach the required precision to detect the GR effects (Eisenhauer et al. 2011). GRAVITY was designed and built in the next decade by a French-German-Portuguese Consortium of 6 Institutes (plus ESO), under the PI-ship of Frank Eisenhauer at MPE[5], and installed on Paranal in July 2015. A detailed discussion of this complex and challenging instrument is given in GRAVITY collaboration et al. (2017), and several other publications.

---

[5] https://www.mpe.mpg.de/938240/Overview, https://www.eso.org/public/teles-instr/paranal-observatory/vlt/vlt-instr/gravity/, https://www.eso.org/sci/facilities/paranal/instruments/gravity.html

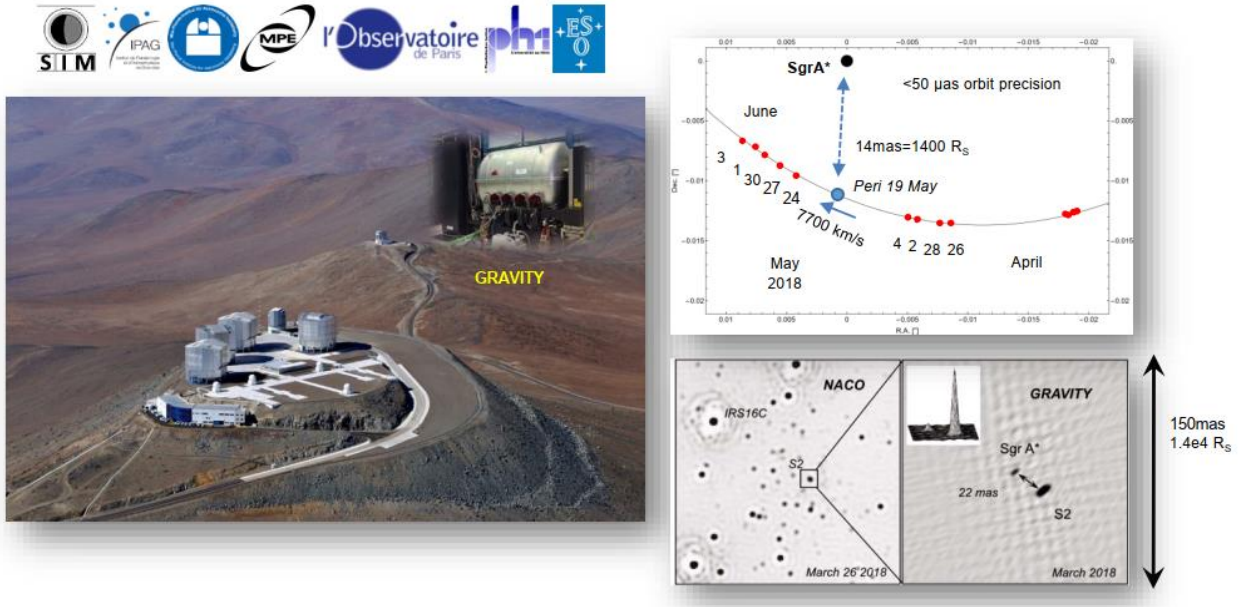

Figure 5. Left: The ESO-Very Large Telescope (VLT) on Cerro Paranal (Chile), where most of the observations by our group were obtained. The Observatory in the Atacama desert is at 2635 m altitude and $-24.7^0$ latitude. It hosts $4\times8.2$m telescopes (large silvered structures), as well as $4\times1.8$m Auxiliary Telescopes (white round domes). Both arrays can be combined optically as a spatial interferometer (VLTI) through mirror trains, where the relative geometric path lengths to a given celestial source can be compensated by movable delay line mirrors in the linear white structure underneath the platform (Glindemann et al. 2003). The final combined set of four beams finally arrives at the beam combiner facility structure underneath the rectangular building in the center of the array. Here the light beams are brought together in the cryogenic beam combiner instrument GRAVITY (built by a French-German-Portuguese consortium of 6 Institutions (logos above the VLT image), plus ESO itself). In GRAVITY we calibrate and optimize the data and extract the visibilities and relative phases of the science object, as well as that of a nearby, fringe tracking reference object, as a function of wavelength, guiding and manipulating the infrared light in single-mode fibers and combining the 6 two-telescope combinations in a micro-chip (GRAVITY collaboration et al. 2017). Bottom right: After calibration of the phases using laser metrology, images with $2\times4$ mas FWHM resolution are reconstructed by Fourier transformation. In the case shown the VLTI science fibers were placed on the star S2/SgrA* in the left image, while the interferometer phases were tracked on the bright star IRS16C 1" NE of SgrA*, in the top left of the AO image. All four telescopes are equipped with infrared adaptive optics, which uses the K=7 bright star IRS7 5" north of S2/SgrA* as a natural guide star to flatten the wavefronts. The image in the bottom right was taken in March 2018, about 2 months before the peri-passage of S2, and both S2 and SgrA* can be clearly detected and its ~22 mas separation measured to ~40-100 μas precision. Top right: During the peri-passage in 2018, the motion of S2 can be easily detected night for night, then moving at ~7700 km/s at ~ 1400 Schwarzschild radii from SgrA* (adapted from Figure 2 of Gravity collaboration et al. 2018a).

Our goal and hope were that the combination of SINFONI, NACO and GRAVITY data would allow us to turn the problem around and use SgrA* as a '*laboratory*' to test General Relativity and the MBH paradigm in a hitherto unexplored regime (e.g. Johannsen 2016). As already mentioned the peri-passage of S2 in May 2018 is a unique opportunity to test GR to PPN1 (e.g. Zucker et al. 2006). Waisberg et al. (2018) showed that a star with a peri-passage 3-5 times smaller than that of S2 may be used to measure the MBH spin through the Lense-Thirring precession of its orbit. Finally, SgrA* itself exhibits continuous variability (Baganoff et al. 2001, Genzel et al. 2003b,

Dodds-Eden et al. 2011, Witzel et al. 2018), and in some cases the fluxes of these 'flares' approach the flux of S2 (K~14), such that 20μas-astrometry on time scales of a few minutes becomes feasible. Several authors had previously speculated that such flares might come from strongly magnetized 'hot spots' of accelerated electrons whose orbital motions might be detectable and used for exploring the innermost accretion zone on the scale of the innermost stable circular orbit, ISCO ($R_{ISCO}$< 6 $R_S$, Broderick & Loeb 2006, Genzel, Eisenhauer & Gillessen 2010, GRAVITY collaboration et al. 2020c,2021).

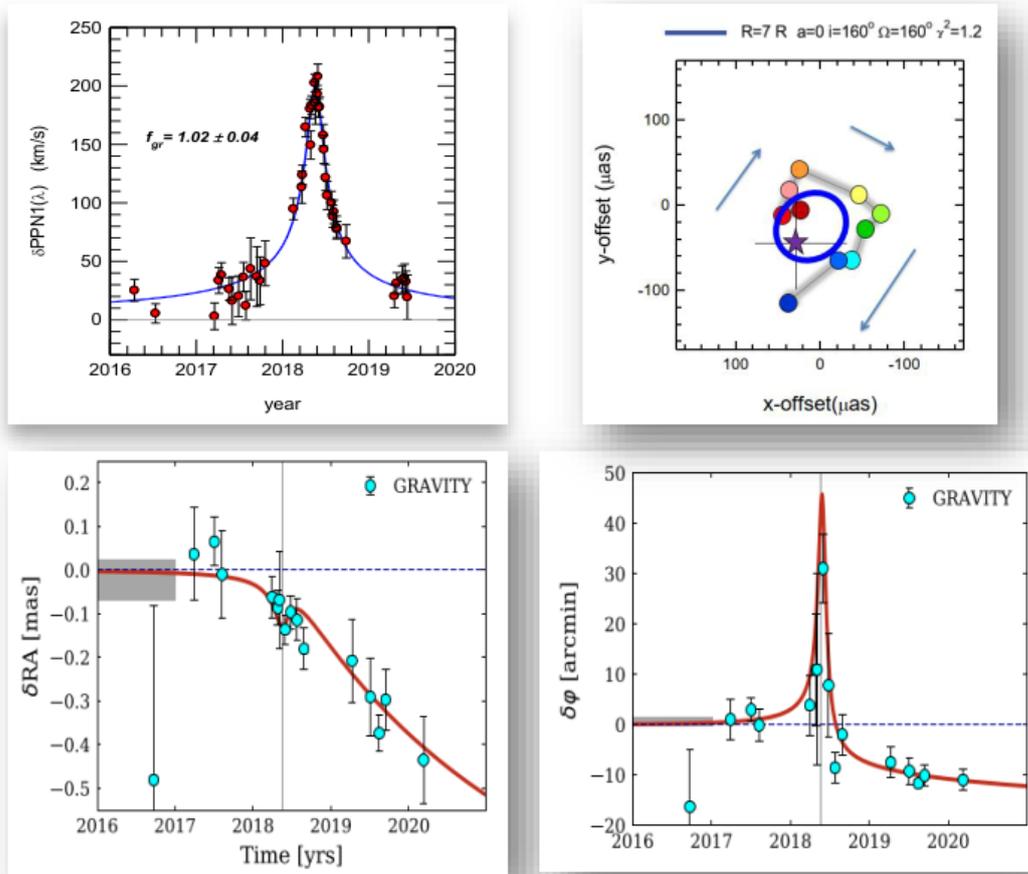

Figure 6. Testing GR and the MBH paradigm with relativistic effects near SgrA*. Top left: Residuals between the SINFONI HeI/HI Brγ line centroid velocities in the local standard of rest (filled red circles with 1σ uncertainties) and the best fitting Newton/Kepler orbit of all spectroscopic and astrometric data over the past three decades (grey horizontal line at 0). The blue line is the best fitting relativistic orbit including all PPN1 terms (as well as Rømer effect), and fitting a free parameter $f_{gr}$ to the PPN1 wavelength term including gravitational redshift and transverse Doppler effect. GR has $f_{gr}$=1 and our best fit yields $f_{gr}$=1.02±0.04 (GRAVITY collaboration et al. 2018a, 2019a, 2020a). Bottom plots: Residuals in RA (left) and angle on the sky φ (right) between the GRAVITY (filled cyan circles and 1σ uncertainties) and average NACO astrometry before 2017 (grey bar) and the best fitting relativistic orbit without precession ($f_{SP}$ =0, blue dotted horizontal line at 0). The best fitting relativistic orbit including precession has $f_{SP}$ =1.1±0.19 (adapted from Figures 3 and B2 in GRAVITY collaboration et al 2020a). Top right: Residual motion of the 2μm light centroid of SgrA* (originating from polarized synchrotron emission from γ>1000 accelerated electrons in the inner accretion zone in a bright 'flare' on July 22$^{nd}$, 2018, cf. Genzel, Eisenhauer &

Gillessen 2010) as a function of time over about 30 minutes, and relative to the location of the mass as estimated from the S2 orbit (dark grey asterisk and 1σ errors). The blue curve denotes a circular particle orbit at 3.5 $R_S$ around a non-spinning MBH of 4.3 million solar masses, inclined at $160^0$ (Figure 1 in GRAVITY collaboration et al. 2018b, 2020b,c, 2021).

It is remarkable to look back in late 2020, two and half years after the peri of S2 on May 19, 2018 and realize that most of these hopes actually turned into reality (Figure 6). The gravitational redshift of S2 has been well determined (5-50σ) by both groups (GRAVITY collaboration et al. 2018a, 2019a, 2020a, Do et al. 2019). The Schwarzschild precession has been detected at ~5σ (GRAVITY collaboration et al. 2020a). Flare motions in three flares of 2018 were consistent with the orbital motions near ISCO around a 4 million solar mass MBH (GRAVITY collaboration et al. 2018b, 2020b). Using the HeI and HI lines as independent 'clocks', GRAVITY collaboration et al. (2019b) have confirmed the local positional invariance of Einstein's equivalence principle to about 5%. Significant upper limits can be placed on the presence of a hypothetical 'fifth force' (Hees et al. 2017, GRAVITY collaboration et al. 2020a). Faint stars close to SgrA* have also been recently detected (GRAVITY collaboration et al. 2021) but are likely not inside the S2 orbit. Overall these discoveries have strengthened the MBH paradigm and GR yet significantly further (Figure 7).

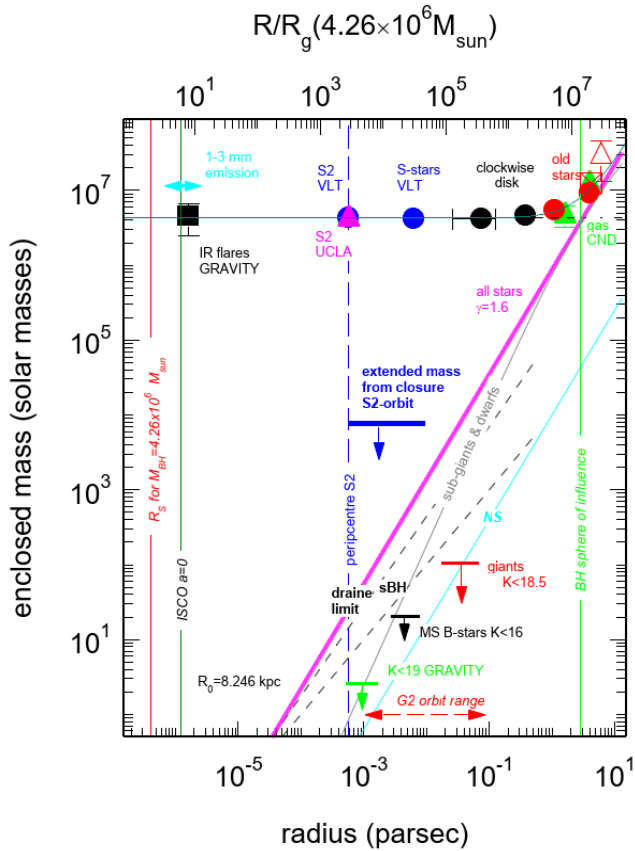

Figure 7. Status of the Galactic Center mass distribution after phase 4. Constraints on the enclosed mass in the central 10 pc of the Galaxy. The blue, black and red circles, the pink, green and red triangles are estimates of the enclosed mass at different radii obtained from stellar and gas motions (see Genzel et al. 2010, GRAVITY collaboration 2021a for details). The filled black rectangle comes from the clockwise loop-motions of synchrotron near-infrared flares (Gravity Collaboration et al. 2018b). The cyan double arrow denotes current VLBI estimates of the 3mm size of Sgr A* (Issaoun et al. 2019). The continuous magenta line shows the total mass model from all stars and stellar remnants (Alexander 2017). The grey line mark the distribution of K < 18.5 sub-giants and dwarfs from Schödel et al. (2018). The grey dashed lines indicate the distribution of stellar black holes and neutron stars from theoretical simulations of Alexander (2017) and Baumgardt et al. (2018), which span a range of roughly a factor 5. Red, black and green upper limits denote upper limits on giants, main-sequence B stars and K < 19 GRAVITY sources. The Schwarzschild radius of a $4.26 \times 10^6$ M$_\odot$ black hole and the innermost stable circular orbit radius for a non-spinning black hole are given by orange and dark green vertical lines. The peri-center radius of S2 is the dashed vertical blue line and the sphere of influence of the black hole is given by the vertical light green line. The blue horizontal line denotes the 2σ upper limit of any extended mass around Sgr A* obtained from the lack of retrograde precession in the S2 orbit (adapted from Figure D1 in Gravity collaboration 2020a).

## 7. Coda

Besides its role at the center stage of testing the black hole paradigm, the Galactic Center has also provided many important ***discoveries and surprises on the astrophysics side***, which I have not described in this paper so far. One is the fact that the central parsec contains ~200 massive, early type stars (O/B and Wolf-Rayet stars), which must have formed in the last few million years (c.f. Sanders 1998, Genzel et al. 1996, 2000, 2010, Paumard et al. 2006, Lu et al. 2009, Bartko et al. 2009). This 'paradox of youth' (Ghez et al. 2003) is completely unexpected, as the MBH should disrupt moderately dense gas clouds tidally, and prevent star formation through local gravitational instabilities and cloud collapse. Perhaps the most likely solution of this riddle is that a large gas cloud fell in a few million years ago, was initially tidally disrupted and shocked, but then cooled and became denser over time, so that gravitational collapse did become possible (cf. Morris & Serabyn 1996, Bonnell & Rice 2008, Hobbs & Nayakshin 2009, Genzel et al. 2010, Alexander 2017).

Possibly connected is the question how the 'S-stars' were captured so close to the MBH, on solar system scales. These B, A, G and K stars could have never migrated in their lifetime to their current position through normal two-body relaxation processes, which take several Gyrs. Instead, rapid stochastic injection of binaries into 'loss-cone' radial orbits from large distances (Hills 1988), and perhaps assisted by massive perturbers (Perets et al. 2007), could have led to a capture of one member of the binary near peri-center, and rapid ejection of the second as a hyper-velocity star (cf. Alexander 2005, 2017, Genzel et al. 2010).

A tidal disruption of a star by the MBH is expected to occur only once every 30,000 years (Alexander 2005, 2017). In 2012 Gillessen et al. (2012, 2019, and references therein) reported the near-radial infall, tidal disruption and eventual slowing down by drag forces near ~2000 $R_S$ of an ionized gas cloud ('G2'). The discussion is ongoing whether this gas cloud is isolated, or whether the gas is the envelope of a central single or binary star.

A third riddle is the lack of a strong cusp of old late type stars around the MBH (Do et al. 2009, Buchholz et al. 2009, Schödel et al. 2018, Figure 7), which is expected in equilibrium (ρ~$R^{-1.5...-1.75}$, Bahcall & Wolf 1977, Alexander 2005, 2017). Finally, the lack of any substantial mass close to SgrA*-MBH greater than a few hundred to one thousand solar masses (Figure 7, GRAVITY collaboration et al. 2020a) is highly exciting and important for other MBH systems and needs to be confirmed by further measurements.

Another aspect I did not cover is the important role MBHs apparently had in the cosmological ***co-evolution with their galactic hosts*** (e.g. Fabian 2012, Kormendy & Ho 2013, Madau & Dickinson 2014).

I have tried to describe in this paper the stepwise progress in proving that massive black holes do exist in the Universe. As compared to the first phase forty years ago, these measurements have pushed the 'size' of the 4 million solar mass concentration downward by almost $10^6$, and its density up by $10^{18}$! Looking ahead toward the future, the question is probably no longer whether SgrA* must be a MBH, but rather whether GR is correct on the scales of the event horizon, whether space-time is described by the Kerr metric and whether the 'no hair theorem' holds. Further improvements of GRAVITY (to GRAVITY$^+$) and the next generation 25-40m telescopes (the ESO-ELT, the TMT and the GMT) promise further progress. A test of the no hair theorem in the Galactic Center might come from combining the stellar dynamics with EHT measurements of the photon ring of SgrA* (Falcke, Melia & Algol 2000, Psaltis & Johannsen 2011, Psaltis, Wex & Kramer 2016, Johannsen 2016). The gravitational waves emanating from the extreme mass ratio in-spiral of a stellar black hole into a massive black hole with the LISA space mission[6] might provide the ultimate culmination of this exciting journey, which Albert Einstein started more than a century ago.

*Acknowledgements: I would like to thank Odele Straub, Tim de Zeeuw, Frank Eisenhauer, Stefan Gillessen, Hannelore Hämmerle, Luis Ho, Pierre Léna, Alvio Renzini, Luciano Rezzolla, Linda Tacconi, Scott Tremaine & Hannah Übler for substantial help with, and comments on this manuscript. I have tried to describe in this paper the journey my colleagues and I took for the past 40 years. The road was long, took patience and hard work, but was enormously rewarding. I am deeply grateful to the many outstanding colleagues who have been willing to work with me on this project, and to the Max-Planck Gesellschaft and the European Southern Observatory for supporting us.*

# References

Abbott, B.P. et al. 2016 Ph.Rev.L. 116, 1102
Abbott, B.P. et al. 2016 Ph.Rev.L. 116, 1101
Alexander, T. 2005, Ph.Rev. 419, 65
Alexander, T. 2017, ARA&A 55, 17
Almheiri, A. Marolf, D., Polchinski, J. & Sully, J. 2013. JHEP 02. 62
Argon, A.L., Greenhill, L.J., Reid, M.J., Moran, J.M. & Humphreys, E.M.L. 2007, ApJ, 659, 1040
Backer, D. C., Zensus, J. A., Kellermann, K. I., Reid, M., Moran, J. M. & Lo, K. Y. 1993, Science 262, 1414
Baganoff, F. et al.2001, Nature, 413, 45
Bahcall, J. N. & Wolf, R.A. 1977, ApJ 216, 883
Balick, B. & Brown, R.L. 1974, ApJ 194, 265
Bardeen J. M., 1973, in Black holes (Les astres occlus), DeWitt B. S., DeWitt C., eds., New York: Gordon and Breach, p. 215
Bartko, H. et al. 2009, ApJ 697, 1741
Baumgardt, H., Amaro-Seoane, P. & Schödel ,R. 2018 A&A 609, 28
Becklin, E.E. & Neugebauer, G. 1968, ApJ 151, 145
Blandford, R.D.  1999, ASPC, 160, 265
Blandford, R.D. , Meier, D. & Readhead, A. 2019, ARA&A 57, 467
Bonaccini-Calia, D., et al. 2006, SPIE 6272, 207
Boehle, A., Ghez, A.M., Schödel, R. et al. 2016, ApJ 830, 17
Bonnell, I.A. & Rice, W.K.M. 2008, Science 321, 1060
Bousso, R. 2002, Rev.Mod.Phys. 74, 825
Bower, G.C. et al. 2004, Science 304, 704
Bower, G. C., Goss, W. M., Falcke, H., Backer, D. C. & Lithwick, Y. 2006, ApJ 648, L127
Broderick, A. & Loeb, A. 2006, MNRAS 367, 905
Broderick, A., Loeb, A. & Narayan, R. 2009, ApJ, 701, 1357
Buchholz, R.M. et al. 2009, A&A 499, 483
Chatterjee, P., Hernquist, L. & Loeb, A. 2002, ApJ, 572, 371

---

[6] https://www.elisascience.org/

Chatzopoulos, S., Fritz, T., Gerhard, O., Gillessen, S., Wegg, C., Genzel, R. & Pfuhl, O. 2015, MNRAS 447, 948
Christou, J.C. 1991, PASP 103, 1040
Crawford, M.K., Genzel,R. Harris, A.I.,Jaffe, D. T., Lacy, J. H., Lugten, J. B., Serabyn, E. & Townes, C. H. 1985, Nature 315, 467
Do, T., Ghez, A. M., Morris, M. R., Yelda, S., Meyer, L., Lu, J. R., Hornstein, S. D.& Matthews, K. 2009, ApJ, 691, 1021
Do, T. et al. 2013, ApJ 779, L6
Do, T. et al. 2018, ApJ 855, L5
Do, T. et al. 2019, Science 365, 664
Dodds-Eden, K. et al. 2011, ApJ, 728, 37
Doeleman, S.S. et al. 2008, Nature, 455, 78
Doeleman, S.S. 2010, in Proceedings of the 10th European VLBI Network Symposium and EVN Users Meeting: VLBI and the new generation of radio arrays. September 20-24, 2010. Manchester, UK. Published online at http://pos.sissa.it/cgi-bin/reader/conf.cgi?confid=125, id.53
Eckart, A. & Genzel, R. 1996, Nature 383, 415
Eckart, A. & Genzel, R. 1997, MNRAS, 284, 576
Eckart, A., Genzel, R., Ott, T. & Schödel, R. 2002, MNRAS 331, 917
Eckart, A. et al., 2006, A&A 450, 535
Einstein, A. 1916, Ann.Phys. 49, 50
Eisenhauer, F. et al. 2003a, ApJ 597, L121
Eisenhauer, F., Tecza, M., Thatte, N. et al. 2003b, ESO Messenger 113, 17
Eisenhauer, F. et al. 2005, ApJ 628, 246
Eisenhauer et al. 2008, in "The Power of Optical/IR Interferometry: Recent Scientific Results and 2nd Generation Instrumentation", ESO Astrophysics Symposia. ISBN 978-3-540-74253-1. Springer, 2008, p. 431
Eisenhauer, F. et al. 2011, ESO Msngr.143, 16
Fabian, A.C., Iwasawa, K., Reynolds, C.S. & Young, A.J. PASP 112, 1145
Fabian, A.C. 2012, ARA&A 50,455
Falcke, H., Melia, F. & Algol, E. 2000, ApJ 528, L13
Feldmeier, A. et al. 2014, A&A 570, 2
Fritz, T.K. et al. 2016, ApJ 821, 44
Genzel, R. & Townes, C.H. 1987, ARA&A 25, 377
Genzel, R., Hollenbach, D., & Townes, C. H., 1994, Rep. Prog. Phys., 57, 417
Genzel, R., Thatte, N., Krabbe, A., Kroker, H. & Tacconi-Garman, L.E. 1996, ApJ, 472, 153
Genzel, R., Eckart, A., Ott, T. & Eisenhauer, F. 1997, MNRAS, 291, 219
Genzel, R., Pichon, C., Eckart, A., Gerhard, O.E. & Ott, T. 2000, MNRAS, 317, 348
Genzel, R. et al. 2003a, ApJ 594, 633
Genzel, R. et al. 2003b, Nature, 425, 934
Genzel, R., Eisenhauer, F. & Gillessen, S. 2010, Rev.Mod.Phys. 82, 3121
Ghez, A.M., Klein, B.L., Morris, M. & Becklin, E.E. 1998, ApJ 509, 678
Ghez, A.M. et al. 2000, Nature, 407, 349
Ghez, A.M. et al. 2003, ApJ 586, L127
Ghez, A.M. et al. 2005, ApJ, 620, 744
Ghez, A.M. et al. 2008, ApJ, 689, 1044
Giacconi, R., Gursky, H., Paolini, F. & Rossi, B.B. 1962, Phys.Rev.Lett. 9, 439
Giacconi, R. 2003, Rev.Mod.Phys. 75, 995
Gillessen, S. et al. 2009a, ApJ, 692, 1075
Gillessen, S. et al. 2009b, ApJ 707, L114
Gillessen, S. et al. 2012, Nature 481, 51
Gillessen, S. et al. 2017, ApJ 837, 30
Glindeman, A. et al. 2003, Astroph.Sp.Sci. 286, 35

Goddi, C. Falcke, H., Kramer, M. et al. 2017, IJMPD, 2630001
Gravity Collaboration et al. 2017, A&A 602, 94
Gravity Collaboration et al. 2018a, A&A 615, L15
Gravity Collaboration et al. 2018b, A&A 618, L10
Gravity Collaboration et al. 2019a, A&A 625, L10
Gravity Collaboration et al. 2019b, Phys.Rev.Lett. 122, 1102
Gravity Collaboration et al. 2020a, A&A 636, L5
Gravity Collaboration et al. 2020b, A&A 635, 143
Gravity Collaboration et al. 2020c, A&A 643, 56
Gravity Collaboration et al. 2021, A&A in press (ArXiv201103058)
Greene, J. et al. 2016, ApJ 826, L32
Gültekin, K. et al. 2009, ApJ 698, 198
Habibi, M. et al. 2019, ApJ 872, L15
Haller, J.W., Rieke, M.J., Rieke, G.H., Tamblyn, P., Close, L. & Melia, F. 1996, ApJ, 456, 194
Hees, A. et al. 2017, Ph.RevL. 118, 1101
Hills, J.G. 1988, Nature 331, 687
Hobbs, A. & Nayakshin, S. 2009, MNRAS 392, 191
Hofmann, K.H. & Eigelt, H. 1993, A&A 278, 328
Issaoun, S. et al. 2019, ApJ 871, 30
Jia, S., Lu, J.R., Sakai, S. et al. 2019, ApJ 873,9
Johannsen, T. 2016, CQGra 33,3001
Johnson, M.D. et al. 2015, Science 350, 1242
Kerr, R.1963, Ph.Rev.Lett., 11, 237
Kormendy, J. 2004, in 'Coevolution of Black Holes and Galaxies', Carnegie Observatories
   Centennial Symposia. Cambridge University Press, Ed. L.C. Ho, p. 1
Kormendy, J. & Ho, L. 2013, ARAA 51, 511
Krabbe, A., Genzel, R., Drapatz, S. & Rotaciuc, V. 1991, ApJ 382, L81
Krabbe, A. et al. 1995, ApJ, 447, L95
Krichbaum, T.P. et al. 1993, A&A 274, 37
Lacy, J.H., Townes, C.H., Geballe, T.R. & Hollenbach, D.J. 1980, ApJ 241, 132
Laplace, S.P. 1795, Le Systeme de Monde, Vol.II. Paris
Larkin, J., Barczys, M., Krabbe, A. et al. 2006, New AR 50, 362
Lena, P. 1991, Science 251, 854
Lenzen, R. & Hofmann, R. 1995, SPIE 2475, 268
Lenzen, R. et al. 2003, SPIE 4841, 944
Liebling, S.L. & Palenzuela, C. 2012, LRR, 15, 6
Lo, K.Y. Backer, D.C., Ekers, R.D. et al. 1985, Nature 315, 124
Lynden-Bell, D. 1969, Nature 223, 690
Lynden-Bell, D. & Rees, M. 1971, MNRAS 152, 461
Lu, J.R. et al. 2009, ApJ 690, 1463
Lu, R.-S. et al. 2014, ApJ 788, L120
Lu, R.-S. et al. 2018, ApJ 859, 60
Madau, P. & Dickinson, M. 2014, ARA&A 52, 415
Magorrian, J. et al. 1998, AJ, 115, 2285
Maldacena, J. 1998, Ad. Th.Math.Phys. 2, 231
Maoz, E. 1995, ApJ, 447, L91
Maoz, E., 1998, ApJ 494, L181
Matthews, K. & Soifer, B.T. 1994, Exp.Astr. 3, 77
Max, C.E. et al. 1997, Science 277, 1649
Mayer, L., Kazantzidis, S., Escala, A.& Callegari, S. 2010, Nature, 466, 1082
McClintock, J. & R. Remillard 2004, in Compact Stellar X-ray sources , eds. W.Lewin and M.van
   der Klis, Cambirdge Univ. Press (astro-ph 0306123)

McConnell, N. & Ma, C.-P. 2013, ApJ 764, 184
McGinn, M.T., Sellgren, K., Becklin, E.E. & Hall, D.N.B. 1989, ApJ, 338, 824
Melia, F. & Falcke, H. 2001, ARA&A 39, 309
Meyer, L., Ghez, A. M., Schödel, R., Yelda, S., Boehle, A., Lu, J. R., Do, T., Morris, M. R., Becklin, E. E.& Matthews, K. 2012, Sci, 338, 84
Michell, J. 1784, Phil. Trans.Royal Soc.London, 74, 35
Miyoshi, M. et al. 1995, Nature 373, 127
Moran, J.M. 2008, ASPC, 395, 87
Morris, M.R. & Serabyn, E. 1996, ARA&A 34, 645
Morris, M.R., Meyer, L. & Ghez, A.M.2012, RAA 12, 995
Munyaneza, F, Tsiklauri, D. & Viollier, R.D. 1998, ApJ, 509, L105
Nandra, K., George, I.M., Mushotzky, R.F., Turner, T.J. & Yaqoob, T.1997, ApJ 477, 602
Oort, J. 1977, ARA&A 15, 295
Osmer. P.S. 2004, in Coevolution of Black Holes and Galaxies, from the Carnegie Observatories Centennial Symposia. Published by Cambridge University Press, as part of the Carnegie Observatories Astrophysics Series. Edited by L. C. Ho, p. 324.
Özel, F., Psaltis, D., Narayan, R. & McClintock, J. E. 2010, ApJ 725, 1918
Paumard, T. et al. 2006, ApJ 643, 1011
Paumard et al. 2008, in "The Power of Optical/IR Interferometry: Recent Scientific Results and 2nd Generation Instrumentation", ESO Astrophysics Symposia. ISBN 978-3-540-74253-1. Springer, 2008, p. 313
Penrose,R. 1963, Ph.Rev.L. 10, 66
Penrose, R. 1965, Ph.Rev.L. 14, 57
Perets, H. et al. 2007 ApJ 656, 709
Plewa, P.M. et al. 2015, MNRAS 453, 3234
Psaltis,D. & Johanssen, T. 2011, JPhCS, 283, 2030
Psaltis, D., Wex, N. & Kramer, M. 2016, ApJ 818,121
Rabien, S., Ott, T., Hackenberg, W. et al. 1998, ApJ 498, 278
Reid, M.J. & Brunthaler, A. 2004, ApJ 616, 872
Reid, M.J. 2009, IJMPD 18, 889
Reid, M.J. Braatz, J. A., Condon, J. J., Lo, K. Y., Kuo, C. Y., Impellizzeri, C. M. V.& Henkel, C. 2013, ApJ, 767, 154
Reid, M.J. et al. 2014, ApJ, 783, 130
Reid, M.J. & Brunthaler, A. 2020, ApJ 892, 39
Remillard, R.A. & McClintock, J.E. 2006, ARA&A 44, 49
Rezzolla, L. et al. 2018, ApJ 852, L25
Rieke, G.H. & Rieke, M.J. 1988, ApJ, 330, L33
Rousset, G., Fontanella, J.C., Kern, P., Gigan, P., Rigaut, F., Léna, P., Boyer, C., Jagourel, P., Gaffard, J.P. & Merkle, F. 1990, A&A 230, L29
Saglia, R. et al. 2016, ApJ 818, 47
Sakai, S. et al. 2019, ApJ 873, 65
Sanders, R.H. 1998, MNRAS, 294, 35
Schödel, R. et al. 2002, Nature 419, 694
Schödel, R. et al. 2003, ApJ 596, 1015
Schödel, R. et al. 2018 A&A 609, 27
Schmidt, M. 1963, Nature 197, 1040
Schunck, F.E. & Mielke, E.W. 2003, CQW 20, R301
Schwarzschild, K., 1916, Sitzungsber. Preuss. Akad.Wiss., 424
Sellgren, K., McGinn, M.T., Becklin, E.E. & Hall, D.N. 1990, ApJ 359, 112
Serabyn, E. & Lacy, J.H. 1985, ApJ 293, 445
Shakura, N.I. & Sunyaev, R.A. 1973, A&A 24, 337
Shen, Z.-Q., Lo, K.Y., Liang, M.C., Ho, P.T.P. & Zhao, J.H. 2005, Nature 438, 62

Sibille, F. Chelli, A.& Léna, P. 1979 A&A, 79, 315
Susskind, L. 1995, JMP 36, 6377
Tanaka, Y., Nandra, K., Fabian, A.C. et al. 1995, Nature 375, 659
Torres,D.F., Capoziello, S. & Lambiase, G. 2000, PhRv D, 62, 4012
Townes, C. H.; Lacy, J. H.; Geballe, T. R. & Hollenbach, D. J. 1982, Nature 301, 661
Tremaine, S., Kocsis, N. & Loeb A. 2021, ArXiv:2012.13273
Trippe, S. et al. 2008, A&A 492, 419
Tsiklauri,D. & Viollier, R. 1998, ApJ 500, 591
Tyson, R.K. & Wizinowich, P. 1992, Phys.Tod. 45, 100
Yelda, S. et al. 2014, ApJ 783, 131
Yuan, F. & Narayan, R. 2014, ARA&A 52, 529
Viollier, R.D, Trautmann, D. & Tupper 1993, PhLB, 306, 79
Waisberg, I. et al. 2018, MNRAS 476, 3600
Weinberg, N. N., Milosavljevic, M. & Ghez, A. M. 2005, ApJ 622, 878
Weitzel, L., Cameron, M., Drapatz, S., Genzel, R. & Krabbe, A. 1994, Exp.Astr. 3,1
Wheeler, J.A. 1968, Amer.Scient. 56, 1
Will, C.M. 2008, ApJ, 674, L25
Witzel, G. et al. 2018, ApJ 863, 15
Wollman, E. R.; Geballe, T. R.; Lacy, J. H.; Townes, C. H.; Rank, D. M.1977, ApJ 218, L103
Zucker, S., Alexander, T., Gillessen, S., Eisenhauer, F. & Genzel, R. 2006, ApJ 639, L21